\documentclass[12pt]{article}
\usepackage[russian]{babel}
\usepackage{amssymb}
\usepackage{multirow}
\usepackage{enumerate}
\usepackage{hhline}
\usepackage{amsmath}
\usepackage{amsthm}
\usepackage{listings}
\usepackage{graphicx}
\usepackage[table,usenames,dvipsnames]{xcolor}
\usepackage{url}
\usepackage[unicode,bookmarksopen,bookmarksopenlevel=1,bookmarksnumbered]{hyperref}
\hypersetup{pdfstartview={XYZ null null 1.25}}

\newtheorem{theorem}{Theorem}

\newtheorem{example}{Example}

\newtheorem{algorithm}{Algorithm}

\def\hcorrection{\hspace{-0.3em}}

\def\Author#1{\vspace{4.0ex plus 0.2ex minus 0.2ex}\centerline{\Large{#1}}}
\def\Title#1{\section*{\hcorrection{#1}}}

\def\References#1{{\footnotesize\baselineskip=12pt\begin{thebibliography}{99}{#1}\end{thebibliography}}}

\def\sps{, \ \ }
\def\spsd{. \ \ }
\begin{document}
\renewcommand{\proofname}{Proof.}
\renewcommand{\refname}{References}
\renewcommand{\figurename}{Fig.}
\renewcommand{\contentsname}{Contents}
\renewcommand{\tablename}{Table}
\renewcommand{\listtablename}{Tables}
\renewcommand{\listfigurename}{Figures}
\lstset{language=Delphi,basicstyle=\tiny,commentstyle=\color{green}}
\lstset{numbers=left, numberstyle=\tiny, stepnumber=1, numbersep=5pt,formfeed=\newpage}
\begin{center}
\Title{On the classification of metric hypercomplex group alternative-elastic algebras for n=8.}
\Author{K.V. Andreev}
\end{center}

In this article, the clarification to Note 4 \cite{Andreev3} for n=8 is considered. In this connection, answers to the following questions are given.\\
\begin{enumerate}
\item  How to classify the metric hypercomplex orthogonal group alternative-elastic algebras for n=8?
\item  How to associate the metric hypercomplex orthogonal group alternative-elastic algebra to the symmetric controlling spin-tensor for n=8?
\item  How technically to construct the symmetric controlling spin-tensor for n=8?
\end{enumerate}

Let \emph{product} of elements of \emph{hypercomplex group alternative-elastic algebra} $\mathbb A$ \cite{Andreev3},\cite{Albert1},\cite{Moreno1}, \cite{Andreev0}, \cite{Andreev1} over the field $\mathbb R$ and the vector space $\mathbb R^8$ be defined as: 1). $\exists !c=ab$; 2). $\exists !e: ae=ea=a$; 3). $\forall a\ne 0\ \exists ! a^{-1}$: $aa^{-1}=a^{-1}a=e$; 4). $\forall a,b:\ (aa)b-a(ab)=b(aa)-(ba)a$; 5). $\forall a,b:\ a(ba)=(ab)a$; 6). $\forall a,b,c:\ a(b+c)=ab+ac,\ (b+c)a=ba+ca$.
Construct one of such the algebras. Suppose that the vector space $\mathbb R^8$ is equipped with the metric  ($i,j,...,A,B,...=\overline{1,8}$)
\begin{equation}
\begin{array}{c}
<a,b>e:=\frac{1}{2}(a\bar b+b\bar a)\sps a=a_1e+\sum\limits_{r=2}^8a_re_r\sps\bar a=a_1e-\sum\limits_{r=2}^8a_re_r\spsd\\
\end{array}
\end{equation}
Then according to the equation (5) \cite{Andreev3}
\begin{equation}
\eta_{ij}{}^k:=\sqrt{2}\eta_i{}^{AB}\eta_{jCA}\eta^k{}_{DB}\theta^{CD}\sps
\end{equation}
where $\eta_{ij}{}^k$ are the structural constants of the algebra, $\eta_i{}^{AB}$ are the connecting operators: a solution of the Clifford equation (4) \cite{Andreev3}.  According to the equation (48) \cite{Andreev3}
\begin{equation}
\label{e1}
\theta^{CD}:=(\theta_0)^{CD}+(\theta_a)^{CD}\sps
\end{equation}
according to the equation (46) \cite{Andreev3}
\begin{equation}
(\theta_a)^{CD}=(\theta_a)^{DC}:=
-\frac{4}{3\sqrt{2}N}(\eta_a)_{lm}{}^r\eta^l{}_{XY}\eta^m{}^{XC}\eta_r{}^{DY}=
\frac{4}{3\sqrt{2}N}(\eta_a)_{lm}{}^r\eta^l{}_{XY}\eta^m{}^{CX}\eta_r{}^{DY}\sps
\end{equation}
according to the equation (44) \cite{Andreev3}
\begin{equation}
\label{e3}
(\theta_0)^{CD}=\frac{2}{N}\varepsilon^{CD}\sps
\end{equation}
where $\varepsilon^{CD}$ is the metric spin-tensor on the spinor space, and for $n=8,\ N=8$. Let the connecting operators for n=8 be given according to \cite[eq. (439), p. 66(eng), eq. (439), p. 177(88)(rus)]{Andreev2} and the matrixes of the metric tensors have the form \cite[eq. (438), p. 65(eng), eq. (438), p. 177(88)(rus)]{Andreev2}
\begin{equation}
    \begin{array}{llll}
    \eta^2{}_{12}=-\frac{1}{\sqrt{2}}\sps  &
    \eta^2{}_{34}=-\frac{1}{\sqrt{2}}\sps  &
    \eta^4{}_{12}=+\frac{i}{\sqrt{2}}\sps  &
    \eta^4{}_{34}=-\frac{i}{\sqrt{2}}\sps  \\
    \eta^2{}_{78}=-\frac{1}{\sqrt{2}}\sps  &
    \eta^2{}_{56}=-\frac{1}{\sqrt{2}}\sps  &
    \eta^4{}_{78}=+\frac{i}{\sqrt{2}}\sps  &
    \eta^4{}_{56}=-\frac{i}{\sqrt{2}}\sps  \\
    \eta^5{}_{14}=+\frac{i}{\sqrt{2}}\sps  &
    \eta^5{}_{23}=-\frac{i}{\sqrt{2}}\sps  &
    \eta^7{}_{14}=-\frac{1}{\sqrt{2}}\sps  &
    \eta^7{}_{23}=-\frac{1}{\sqrt{2}}\sps  \\
    \eta^5{}_{67}=+\frac{i}{\sqrt{2}}\sps  &
    \eta^5{}_{58}=-\frac{i}{\sqrt{2}}\sps  &
    \eta^7{}_{67}=-\frac{1}{\sqrt{2}}\sps  &
    \eta^7{}_{58}=-\frac{1}{\sqrt{2}}\sps  \\
    \eta^6{}_{13}=-\frac{i}{\sqrt{2}}\sps  &
    \eta^6{}_{24}=-\frac{i}{\sqrt{2}}\sps  &
    \eta^8{}_{13}=+\frac{1}{\sqrt{2}}\sps  &
    \eta^8{}_{24}=-\frac{1}{\sqrt{2}}\sps  \\
    \eta^6{}_{68}=+\frac{i}{\sqrt{2}}\sps  &
    \eta^6{}_{57}=+\frac{i}{\sqrt{2}}\sps  &
    \eta^8{}_{68}=-\frac{1}{\sqrt{2}}\sps  &
    \eta^8{}_{57}=+\frac{1}{\sqrt{2}}\sps  \\
    \eta^1{}_{15}=+\frac{1}{\sqrt{2}}\sps  &
    \eta^1{}_{51}=+\frac{1}{\sqrt{2}}\sps  &
    \eta^3{}_{15}=-\frac{i}{\sqrt{2}}\sps  &
    \eta^3{}_{51}=+\frac{i}{\sqrt{2}}\sps  \\
    \eta^1{}_{26}=+\frac{1}{\sqrt{2}}\sps  &
    \eta^1{}_{62}=+\frac{1}{\sqrt{2}}\sps  &
    \eta^3{}_{26}=-\frac{i}{\sqrt{2}}\sps  &
    \eta^3{}_{62}=+\frac{i}{\sqrt{2}}\sps  \\
    \eta^1{}_{37}=+\frac{1}{\sqrt{2}}\sps  &
    \eta^1{}_{73}=+\frac{1}{\sqrt{2}}\sps  &
    \eta^3{}_{37}=-\frac{i}{\sqrt{2}}\sps  &
    \eta^3{}_{73}=+\frac{i}{\sqrt{2}}\sps  \\
    \eta^1{}_{48}=+\frac{1}{\sqrt{2}}\sps  &
    \eta^1{}_{84}=+\frac{1}{\sqrt{2}}\sps  &
    \eta^3{}_{48}=-\frac{i}{\sqrt{2}}\sps  &
    \eta^3{}_{84}=+\frac{i}{\sqrt{2}}\spsd
    \end{array}
\end{equation}
We need to define the transformation
\begin{equation}
\label{e2}
\frac{1}{\sqrt{2}}\left(
\begin{array}{cc}
  E & E\\
-iE & iE
\end{array}
\right)
\end{equation}
of the spinor basis, where E is the identity matrix $4\times 4$. In the new basis, the matrix of the metric spin-tensor will have the diagonal form with $\ll+1\gg$ on the main diagonal. The matrix of the involution will have the same form. With the help of such the connecting operators, the double covering $Spin(8,\mathbb R)/\pm1\cong SO(8,\mathbb R)$ is described as
\begin{equation}
S_i{}^j\eta_j{}^{AB}=\eta_i{}^{CD}\tilde S_C{}^A\tilde{\tilde S}_D{}^B\sps
\tilde S_C{}^A\tilde S_D{}^B\varepsilon_{AB}=\varepsilon_{CD}\sps
\tilde{\tilde S}_C{}^A\tilde{\tilde S}_D{}^B\varepsilon_{AB}=\varepsilon_{CD}\sps
\end{equation}
according to Corollary 8.3 \cite[p. 44 (eng), p. 174(51) (rus)]{Andreev1}. If $S_i{}^j$ keeps the algebra identity then $S_C{}^A=\tilde S_C{}^A=\tilde{\tilde S}_C{}^A$.
\begin{example}
\label{ex3}
\texttt{Octonion \cite{Baez1}.}
\begin{table}[h]
\caption{The canonical octonion multiplication table.}
\label{t4}
\begin{center}\footnotesize
\setlength{\tabcolsep}{1pt}
\begin{tabular}{|>{\columncolor{yellow}}c|>{\columncolor[gray]{0.95}}c|>{\columncolor[gray]{0.95}}c|>{\columncolor[gray]{0.95}}c|>{\columncolor[gray]{0.95}}c
|>{\columncolor[gray]{0.95}}c|>{\columncolor[gray]{0.95}}c|>{\columncolor[gray]{0.95}}c|>{\columncolor[gray]{0.95}}c|>{\columncolor[gray]{0.95}}c
|>{\columncolor[gray]{0.95}}c|>{\columncolor[gray]{0.95}}c|>{\columncolor[gray]{0.95}}c|>{\columncolor[gray]{0.95}}c|>{\columncolor[gray]{0.95}}c
|>{\columncolor[gray]{0.95}}c|>{\columncolor[gray]{0.95}}c|}\hline
\rowcolor{yellow} \cellcolor[gray]{0.95}
       *&   $e_{0}$ &    $e_{1}$ &    $e_{2}$ &    $e_{3}$ &    $e_{4}$ &    $e_{5}$ &    $e_{6}$ &    $e_{7}$  \\ \hline
 $e_{0}$&   $e_{0}$ &    $e_{1}$ &    $e_{2}$ &    $e_{3}$ &    $e_{4}$ &    $e_{5}$ &    $e_{6}$ &    $e_{7}$  \\ \hline
 $e_{1}$&   $e_{1}$ &   -$e_{0}$ &    $e_{3}$ &   -$e_{2}$ &    $e_{5}$ &   -$e_{4}$ &   -$e_{7}$ &    $e_{6}$  \\ \hline
 $e_{2}$&   $e_{2}$ &   -$e_{3}$ &   -$e_{0}$ &    $e_{1}$ &    $e_{6}$ &    $e_{7}$ &   -$e_{4}$ &   -$e_{5}$  \\ \hline
 $e_{3}$&   $e_{3}$ &    $e_{2}$ &   -$e_{1}$ &   -$e_{0}$ &    $e_{7}$ &   -$e_{6}$ &    $e_{5}$ &   -$e_{4}$  \\ \hline
 $e_{4}$&   $e_{4}$ &   -$e_{5}$ &   -$e_{6}$ &   -$e_{7}$ &   -$e_{0}$ &    $e_{1}$ &    $e_{2}$ &    $e_{3}$  \\ \hline
 $e_{5}$&   $e_{5}$ &    $e_{4}$ &   -$e_{7}$ &    $e_{6}$ &   -$e_{1}$ &   -$e_{0}$ &   -$e_{3}$ &    $e_{2}$  \\ \hline
 $e_{6}$&   $e_{6}$ &    $e_{7}$ &    $e_{4}$ &   -$e_{5}$ &   -$e_{2}$ &    $e_{3}$ &   -$e_{0}$ &   -$e_{1}$  \\ \hline
 $e_{7}$&   $e_{7}$ &   -$e_{6}$ &    $e_{5}$ &    $e_{4}$ &   -$e_{3}$ &   -$e_{2}$ &    $e_{1}$ &   -$e_{0}$  \\ \hline
\end{tabular}
\end{center}
\end{table}
\\Then in the new basis, the matrix of the symmetric spin-tensor $\theta^{CD}$ (and $\theta_{CD}$, because the matrix of $\varepsilon_{CD}$ is the identity one) has the form
\begin{equation}
\left(
\begin{array}{rrrrrrrr}
 2 & 0 & 0 & 0 & 0 & 0 & 0 & 0 \\
 0 & 0 & 0 & 0 & 0 & 0 & 0 & 0 \\
 0 & 0 & 0 & 0 & 0 & 0 & 0 & 0 \\
 0 & 0 & 0 & 0 & 0 & 0 & 0 & 0 \\
 0 & 0 & 0 & 0 & 0 & 0 & 0 & 0 \\
 0 & 0 & 0 & 0 & 0 & 0 & 0 & 0 \\
 0 & 0 & 0 & 0 & 0 & 0 & 0 & 0 \\
 0 & 0 & 0 & 0 & 0 & 0 & 0 & 0 \\
\end{array}
\right)\spsd
\end{equation}
In the transition to the subgroup keeping the octonion identity, the dimension reduces on 7, in the transition to the subgroup keeping the symmetric spin-tensor, the dimension reduces on 7 else then the dimension of $Aut(\mathbb A^8)$ is equal to $\frac{8*7}{2}-7-7=14$. Thus, we obtain the Lie group $G_2$. Passing to the infinitesimal transformations according to Section 10 \cite[p. 52 (eng), pp. 183-184(60-61) (rus)]{Andreev1}, we obtain
\begin{equation}
T_C{}^A=\frac{1}{2}T^{ij}\eta_j{}^{AB}\eta_i{}_{CB}\spsd
\end{equation}
The preservation of the metrics means that
\begin{equation}
T_{CA}=-T_{AC}\sps T_{ij}=-T_{ji}\spsd
\end{equation}
The dimension of the antisymmetrical space is equal to $\frac{8*7}{2}=28$. In the transition to the subgroup keeping the octonion identity, we obtain
\begin{equation}
\eta_i{}^{AB}T_{AB}=0\sps\eta^iT_{ij}=0\spsd
\end{equation}
In the new basis,
\begin{equation}
\label{e20}
\left\{
\begin{array}{l}
-T_{12}-T_{34}+T_{56}+T_{78}=0\sps\\
+T_{13}-T_{24}-T_{57}+T_{68}=0\sps\\
-T_{14}-T_{23}+T_{58}+T_{67}=0\sps\\
+T_{15}+T_{26}+T_{37}+T_{48}=0\sps\\
+T_{16}-T_{25}-T_{38}+T_{47}=0\sps\\
-T_{17}-T_{28}+T_{35}+T_{46}=0\sps\\
+T_{18}-T_{27}+T_{36}-T_{45}=0\spsd
\end{array}
\right.
\end{equation}
This reduces the dimension on 7. In the transition to the subgroup keeping the symmetric spin-tensor, we obtain
\begin{equation}
T_A{}^C\theta_{CB}+T_B{}^C\theta_{AC}=0\spsd
\end{equation}
In the new basis,
\begin{equation}
\label{e21}
T_{12}=0\sps T_{13}=0\sps T_{14}=0\sps T_{15}=0\sps T_{16}=0\sps T_{17}=0\sps T_{18}=0\spsd
\end{equation}
Because the symmetric spin-tensor has the single significant eigenvalue, this reduces the dimension on 7 else. Thus, we obtain the Lie algebra $g_2$ whose the dimension is equal to 14. The system (\ref{e20}) with the relations (\ref{e21}) coincides with the one \cite[eq. (13), p. 313]{Postnicov1} up to an orthogonal transformation.
\end{example}

\newpage
\begin{example}
\label{ex4}
\texttt{The generating octonion algebra (for $e_1$).}
\begin{table}[h]
\caption{The multiplication table of the generating octonion algebra.}
\label{t5}
\begin{center}\footnotesize
\setlength{\tabcolsep}{1pt}
\begin{tabular}{|>{\columncolor{yellow}}c|>{\columncolor[gray]{0.95}}c|>{\columncolor[gray]{0.95}}c|>{\columncolor[gray]{0.95}}c|>{\columncolor[gray]{0.95}}c
|>{\columncolor[gray]{0.95}}c|>{\columncolor[gray]{0.95}}c|>{\columncolor[gray]{0.95}}c|>{\columncolor[gray]{0.95}}c|>{\columncolor[gray]{0.95}}c
|>{\columncolor[gray]{0.95}}c|>{\columncolor[gray]{0.95}}c|>{\columncolor[gray]{0.95}}c|>{\columncolor[gray]{0.95}}c|>{\columncolor[gray]{0.95}}c
|>{\columncolor[gray]{0.95}}c|>{\columncolor[gray]{0.95}}c|}\hline
\rowcolor{yellow} \cellcolor[gray]{0.95}
       *&   $e_{0}$ &    $e_{1}$ &    $e_{2}$ &    $e_{3}$ &    $e_{4}$ &    $e_{5}$ &    $e_{6}$ &    $e_{7}$  \\ \hline
 $e_{0}$&   $e_{0}$ &    $e_{1}$ &    $e_{2}$ &    $e_{3}$ &    $e_{4}$ &    $e_{5}$ &    $e_{6}$ &    $e_{7}$  \\ \hline
 $e_{1}$&   $e_{1}$ &   -$e_{0}$ &    $e_{3}$ &   -$e_{2}$ &    $e_{5}$ &   -$e_{4}$ &   -$e_{7}$ &    $e_{6}$  \\ \hline
 $e_{2}$&   $e_{2}$ &   -$e_{3}$ &   -$e_{0}$ &    $e_{1}$ &            &            &            &             \\ \hline
 $e_{3}$&   $e_{3}$ &    $e_{2}$ &   -$e_{1}$ &   -$e_{0}$ &            &            &            &             \\ \hline
 $e_{4}$&   $e_{4}$ &   -$e_{5}$ &            &            &   -$e_{0}$ &    $e_{1}$ &            &             \\ \hline
 $e_{5}$&   $e_{5}$ &    $e_{4}$ &            &            &   -$e_{1}$ &   -$e_{0}$ &            &             \\ \hline
 $e_{6}$&   $e_{6}$ &    $e_{7}$ &            &            &            &            &   -$e_{0}$ &   -$e_{1}$  \\ \hline
 $e_{7}$&   $e_{7}$ &   -$e_{6}$ &            &            &            &            &    $e_{1}$ &   -$e_{0}$  \\ \hline
\end{tabular}
\end{center}
\end{table}
\\Then in the new basis, the matrix of the symmetric spin-tensor $\theta^{CD}$ has the form
\begin{equation}
\left(
\begin{array}{rrrrrrrr}
 1 & 0 & 0 & 0 & 0 & 0 & 0 & 0 \\
 0 & 1 & 0 & 0 & 0 & 0 & 0 & 0 \\
 0 & 0 & 0 & 0 & 0 & 0 & 0 & 0 \\
 0 & 0 & 0 & 0 & 0 & 0 & 0 & 0 \\
 0 & 0 & 0 & 0 & 0 & 0 & 0 & 0 \\
 0 & 0 & 0 & 0 & 0 & 0 & 0 & 0 \\
 0 & 0 & 0 & 0 & 0 & 0 & 0 & 0 \\
 0 & 0 & 0 & 0 & 0 & 0 & 0 & 0 \\
\end{array}
\right)\spsd
\end{equation}
In the new basis,  the relations
\begin{equation}
\label{e22}
T_{23}=0\sps T_{24}=0\sps T_{25}=0\sps T_{26}=0\sps T_{27}=0\sps T_{28}=0
\end{equation}
are added to the ones from (\ref{e21}). Because the symmetric spin-tensor has the two significant eigenvalue, this reduces the dimension on 6 else. However, we need to remove the relation $T_{12}=0$ from (\ref{e21}). Thus, we obtain the Lie algebra whose the dimension is equal to 14-6+1=9.
\end{example}

\begin{example}
\label{ex5}
\texttt{The quaternion algebra analog.}
\begin{table}[h]
\caption{The multiplication table of the quaternion algebra analog.}
\label{t6}
\begin{center}\footnotesize
\setlength{\tabcolsep}{1pt}
\begin{tabular}{|>{\columncolor{yellow}}c|>{\columncolor[gray]{0.95}}c|>{\columncolor[gray]{0.95}}c|>{\columncolor[gray]{0.95}}c|>{\columncolor[gray]{0.95}}c
|>{\columncolor[gray]{0.95}}c|>{\columncolor[gray]{0.95}}c|>{\columncolor[gray]{0.95}}c|>{\columncolor[gray]{0.95}}c|>{\columncolor[gray]{0.95}}c
|>{\columncolor[gray]{0.95}}c|>{\columncolor[gray]{0.95}}c|>{\columncolor[gray]{0.95}}c|>{\columncolor[gray]{0.95}}c|>{\columncolor[gray]{0.95}}c
|>{\columncolor[gray]{0.95}}c|>{\columncolor[gray]{0.95}}c|}\hline
\rowcolor{yellow} \cellcolor[gray]{0.95}
       *&   $e_{0}$ &    $e_{1}$ &    $e_{2}$ &    $e_{3}$ &    $e_{4}$ &    $e_{5}$ &    $e_{6}$ &    $e_{7}$  \\ \hline
 $e_{0}$&   $e_{0}$ &    $e_{1}$ &    $e_{2}$ &    $e_{3}$ &    $e_{4}$ &    $e_{5}$ &    $e_{6}$ &    $e_{7}$  \\ \hline
 $e_{1}$&   $e_{1}$ &   -$e_{0}$ &    $e_{3}$ &   -$e_{2}$ &            &            &            &             \\ \hline
 $e_{2}$&   $e_{2}$ &   -$e_{3}$ &   -$e_{0}$ &    $e_{1}$ &            &            &            &             \\ \hline
 $e_{3}$&   $e_{3}$ &    $e_{2}$ &   -$e_{1}$ &   -$e_{0}$ &            &            &            &             \\ \hline
 $e_{4}$&   $e_{4}$ &            &            &            &   -$e_{0}$ &            &            &             \\ \hline
 $e_{5}$&   $e_{5}$ &            &            &            &            &   -$e_{0}$ &            &             \\ \hline
 $e_{6}$&   $e_{6}$ &            &            &            &            &            &   -$e_{0}$ &             \\ \hline
 $e_{7}$&   $e_{7}$ &            &            &            &            &            &            &   -$e_{0}$  \\ \hline
\end{tabular}
\end{center}
\end{table}
\\Then in the new basis, the matrix of the symmetric spin-tensor $\theta^{CD}$ has the form
\begin{equation}
\left(
\begin{array}{rrrrrrrr}
 0.5 & 0   & 0 & 0 & 0   & 0   & 0 & 0 \\
 0   & 0.5 & 0 & 0 & 0   & 0   & 0 & 0 \\
 0   & 0   & 0 & 0 & 0   & 0   & 0 & 0 \\
 0   & 0   & 0 & 0 & 0   & 0   & 0 & 0 \\
 0   & 0   & 0 & 0 & 0.5 & 0   & 0 & 0 \\
 0   & 0   & 0 & 0 & 0   & 0.5 & 0 & 0 \\
 0   & 0   & 0 & 0 & 0   & 0   & 0 & 0 \\
 0   & 0   & 0 & 0 & 0   & 0   & 0 & 0 \\
\end{array}
\right)\spsd
\end{equation}
In the new basis,  the relations
\begin{equation}
T_{53}=0\sps T_{54}=0\sps T_{56}=0\sps T_{57}=0\sps T_{58}=0
\end{equation}
are added to the ones from (\ref{e21}), (\ref{e22}). Because the symmetric spin-tensor has the four significant eigenvalue, this reduces the dimension on 5 else. Thus, we obtain the Lie algebra whose the dimension is equal to 8-5=3. This means that the group $Aut(\mathbb A^4)$ is isomorphic to $SO(3,\mathbb R)$. Of course, we need to append $SO(4,\mathbb R)$ to obtain the all automorphism group.
\end{example}

\begin{example}
\label{ex6}
\texttt{The carcass for the octonion algebra.}
\begin{table}[h]
\caption{The multiplication table of the octonion carcass.}
\label{t7}
\begin{center}\footnotesize
\setlength{\tabcolsep}{1pt}
\begin{tabular}{|>{\columncolor{yellow}}c|>{\columncolor[gray]{0.95}}c|>{\columncolor[gray]{0.95}}c|>{\columncolor[gray]{0.95}}c|>{\columncolor[gray]{0.95}}c
|>{\columncolor[gray]{0.95}}c|>{\columncolor[gray]{0.95}}c|>{\columncolor[gray]{0.95}}c|>{\columncolor[gray]{0.95}}c|>{\columncolor[gray]{0.95}}c
|>{\columncolor[gray]{0.95}}c|>{\columncolor[gray]{0.95}}c|>{\columncolor[gray]{0.95}}c|>{\columncolor[gray]{0.95}}c|>{\columncolor[gray]{0.95}}c
|>{\columncolor[gray]{0.95}}c|>{\columncolor[gray]{0.95}}c|}\hline
\rowcolor{yellow} \cellcolor[gray]{0.95}
       *&   $e_{0}$ &    $e_{1}$ &    $e_{2}$ &    $e_{3}$ &    $e_{4}$ &    $e_{5}$ &    $e_{6}$ &    $e_{7}$  \\ \hline
 $e_{0}$&       $0$ &        $0$ &        $0$ &        $0$ &        $0$ &        $0$ &        $0$ &        $0$  \\ \hline
 $e_{1}$&       $0$ &        $0$ &    $e_{3}$ &   -$e_{2}$ &        $0$ &        $0$ &   -$e_{7}$ &    $e_{6}$  \\ \hline
 $e_{2}$&       $0$ &   -$e_{3}$ &        $0$ &    $e_{1}$ &        $0$ &    $e_{7}$ &       $0$ &    -$e_{5}$  \\ \hline
 $e_{3}$&       $0$ &    $e_{2}$ &   -$e_{1}$ &        $0$ &        $0$ &   -$e_{6}$ &    $e_{5}$ &        $0$  \\ \hline
 $e_{4}$&       $0$ &        $0$ &        $0$ &        $0$ &        $0$ &        $0$ &        $0$ &        $0$  \\ \hline
 $e_{5}$&       $0$ &        $0$ &   -$e_{7}$ &    $e_{6}$ &        $0$ &        $0$ &   -$e_{3}$ &    $e_{2}$  \\ \hline
 $e_{6}$&       $0$ &    $e_{7}$ &        $0$ &   -$e_{5}$ &        $0$ &    $e_{3}$ &        $0$ &   -$e_{1}$  \\ \hline
 $e_{7}$&       $0$ &   -$e_{6}$ &    $e_{5}$ &        $0$ &        $0$ &   -$e_{2}$ &    $e_{1}$ &        $0$  \\ \hline
\end{tabular}
\end{center}
\end{table}
\\Then in the new basis, the matrix of the symmetric spin-tensor $\theta^{CD}$ has the form
\begin{equation}
\left(
\begin{array}{rrrrrrrr}
  1 & 0 & 0 & 0 & 0 & 0 & 0 & 0 \\
  0 & 0 & 0 & 0 & 0 & 0 & 0 & 0 \\
  0 & 0 & 0 & 0 & 0 & 0 & 0 & 0 \\
  0 & 0 & 0 & 0 & 0 & 0 & 0 & 0 \\
  0 & 0 & 0 & 0 & 0 & 0 & 0 & 0 \\
  0 & 0 & 0 & 0 & 0 & 0 & 0 & 0 \\
  0 & 0 & 0 & 0 & 0 & 0 & 0 & 0 \\
  0 & 0 & 0 & 0 & 0 & 0 & 0 &-1 \\
\end{array}
\right)\spsd
\end{equation}
\end{example}

\newpage
\begin{example}
\label{ex7}
\texttt{The generation octonion algebra (for $e_4$).}
\begin{table}[h]
\caption{The multiplication table of the generating octonion algebra.}
\label{t8}
\begin{center}\footnotesize
\setlength{\tabcolsep}{1pt}
\begin{tabular}{|>{\columncolor{yellow}}c|>{\columncolor[gray]{0.95}}c|>{\columncolor[gray]{0.95}}c|>{\columncolor[gray]{0.95}}c|>{\columncolor[gray]{0.95}}c
|>{\columncolor[gray]{0.95}}c|>{\columncolor[gray]{0.95}}c|>{\columncolor[gray]{0.95}}c|>{\columncolor[gray]{0.95}}c|>{\columncolor[gray]{0.95}}c
|>{\columncolor[gray]{0.95}}c|>{\columncolor[gray]{0.95}}c|>{\columncolor[gray]{0.95}}c|>{\columncolor[gray]{0.95}}c|>{\columncolor[gray]{0.95}}c
|>{\columncolor[gray]{0.95}}c|>{\columncolor[gray]{0.95}}c|}\hline
\rowcolor{yellow} \cellcolor[gray]{0.95}
       *&   $e_{0}$ &    $e_{1}$ &    $e_{2}$ &    $e_{3}$ &    $e_{4}$ &    $e_{5}$ &    $e_{6}$ &    $e_{7}$  \\ \hline
 $e_{0}$&   $e_{0}$ &    $e_{1}$ &    $e_{2}$ &    $e_{3}$ &    $e_{4}$ &    $e_{5}$ &    $e_{6}$ &    $e_{7}$  \\ \hline
 $e_{1}$&   $e_{1}$ &        $0$ &        $0$ &        $0$ &    $e_{5}$ &   -$e_{4}$ &        $0$ &        $0$  \\ \hline
 $e_{2}$&   $e_{2}$ &        $0$ &        $0$ &        $0$ &    $e_{6}$ &        $0$ &   -$e_{4}$ &        $0$ \\ \hline
 $e_{3}$&   $e_{3}$ &        $0$ &        $0$ &        $0$ &    $e_{7}$ &        $0$ &        $0$ &   -$e_{4}$  \\ \hline
 $e_{4}$&   $e_{4}$ &   -$e_{5}$ &   -$e_{6}$ &   -$e_{7}$ &   -$e_{0}$ &    $e_{1}$ &    $e_{2}$ &    $e_{3}$  \\ \hline
 $e_{5}$&   $e_{5}$ &    $e_{4}$ &        $0$ &        $0$ &   -$e_{1}$ &        $0$ &        $0$ &        $0$  \\ \hline
 $e_{6}$&   $e_{6}$ &        $0$ &    $e_{4}$ &        $0$ &   -$e_{2}$ &        $0$ &        $0$ &        $0$  \\ \hline
 $e_{7}$&   $e_{7}$ &        $0$ &        $0$ &    $e_{4}$ &   -$e_{3}$ &        $0$ &        $0$ &        $0$  \\ \hline
\end{tabular}
\end{center}
\end{table}
\\Then in the new basis, the matrix of the symmetric spin-tensor $\theta^{CD}$ has the form
\begin{equation}
\left(
\begin{array}{rrrrrrrr}
  1 & 0 & 0 & 0 & 0 & 0 & 0 & 0 \\
  0 & 0 & 0 & 0 & 0 & 0 & 0 & 0 \\
  0 & 0 & 0 & 0 & 0 & 0 & 0 & 0 \\
  0 & 0 & 0 & 0 & 0 & 0 & 0 & 0 \\
  0 & 0 & 0 & 0 & 0 & 0 & 0 & 0 \\
  0 & 0 & 0 & 0 & 0 & 0 & 0 & 0 \\
  0 & 0 & 0 & 0 & 0 & 0 & 0 & 0 \\
  0 & 0 & 0 & 0 & 0 & 0 & 0 & 1 \\
\end{array}
\right)\spsd
\end{equation}
\end{example}

\begin{example}
\label{ex8}
\texttt{The algebra of the octonion type.}
\begin{table}[h]
\caption{The octonion multiplication table (non-canonical).}
\label{t9}
\begin{center}\footnotesize
\setlength{\tabcolsep}{1pt}
\begin{tabular}{|>{\columncolor{yellow}}c|>{\columncolor[gray]{0.95}}c|>{\columncolor[gray]{0.95}}c|>{\columncolor[gray]{0.95}}c|>{\columncolor[gray]{0.95}}c
|>{\columncolor[gray]{0.95}}c|>{\columncolor[gray]{0.95}}c|>{\columncolor[gray]{0.95}}c|>{\columncolor[gray]{0.95}}c|>{\columncolor[gray]{0.95}}c
|>{\columncolor[gray]{0.95}}c|>{\columncolor[gray]{0.95}}c|>{\columncolor[gray]{0.95}}c|>{\columncolor[gray]{0.95}}c|>{\columncolor[gray]{0.95}}c
|>{\columncolor[gray]{0.95}}c|>{\columncolor[gray]{0.95}}c|}\hline
\rowcolor{yellow} \cellcolor[gray]{0.95}
       *&   $e_{0}$ &    $e_{1}$ &    $e_{2}$ &    $e_{3}$ &    $e_{4}$ &    $e_{5}$ &    $e_{6}$ &    $e_{7}$  \\ \hline
 $e_{0}$&   $e_{0}$ &    $e_{1}$ &    $e_{2}$ &    $e_{3}$ &    $e_{4}$ &    $e_{5}$ &    $e_{6}$ &    $e_{7}$  \\ \hline
 $e_{1}$&   $e_{1}$ &   -$e_{0}$ &    $e_{3}$ &   -$e_{2}$ &    $e_{5}$ &   -$e_{4}$ &   -$e_{7}$ &    $e_{6}$  \\ \hline
 $e_{2}$&   $e_{2}$ &   -$e_{3}$ &   -$e_{0}$ &    $e_{1}$ &    $e_{6}$ &    $e_{7}$ &    $e_{5}$ &   -$e_{4}$  \\ \hline
 $e_{3}$&   $e_{3}$ &    $e_{2}$ &   -$e_{1}$ &   -$e_{0}$ &   -$e_{6}$ &   -$e_{7}$ &    $e_{4}$ &    $e_{5}$  \\ \hline
 $e_{4}$&   $e_{4}$ &   -$e_{5}$ &   -$e_{7}$ &    $e_{6}$ &   -$e_{0}$ &    $e_{1}$ &   -$e_{3}$ &    $e_{2}$  \\ \hline
 $e_{5}$&   $e_{5}$ &    $e_{4}$ &    $e_{6}$ &    $e_{7}$ &   -$e_{1}$ &   -$e_{0}$ &   -$e_{2}$ &   -$e_{3}$  \\ \hline
 $e_{6}$&   $e_{6}$ &    $e_{7}$ &   -$e_{5}$ &   -$e_{4}$ &    $e_{2}$ &    $e_{3}$ &   -$e_{0}$ &   -$e_{1}$  \\ \hline
 $e_{7}$&   $e_{7}$ &   -$e_{6}$ &    $e_{4}$ &   -$e_{5}$ &   -$e_{2}$ &    $e_{3}$ &    $e_{1}$ &   -$e_{0}$  \\ \hline
\end{tabular}
\end{center}
\end{table}
\\Then in the new basis, the matrix of the symmetric spin-tensor $\theta^{CD}$ has the form
\begin{equation}
\left(
\begin{array}{rrrrrrrr}
  1 & 1 & 0 & 0 & 0 & 0 & 0 & 0 \\
  1 & 1 & 0 & 0 & 0 & 0 & 0 & 0 \\
  0 & 0 & 0 & 0 & 0 & 0 & 0 & 0 \\
  0 & 0 & 0 & 0 & 0 & 0 & 0 & 0 \\
  0 & 0 & 0 & 0 & 0 & 0 & 0 & 0 \\
  0 & 0 & 0 & 0 & 0 & 0 & 0 & 0 \\
  0 & 0 & 0 & 0 & 0 & 0 & 0 & 0 \\
  0 & 0 & 0 & 0 & 0 & 0 & 0 & 0 \\
\end{array}
\right)\spsd
\end{equation}
\end{example}

\begin{algorithm}
Thus, in order to compare the two algebras for n=8, it is necessary
\begin{enumerate}
\item to lead the both algebras to the same algebra identity;
\item to solve the Clifford equation, in order to find the connecting operators;
\item to construct the controlling symmetric spin-tensor on the structural constants and the connecting operators for each algebra;
\item to find the eigenvalues of the controlling spin-tensors.
\end{enumerate}
Then
\begin{enumerate}
\item if the eigenvalues coincide then such the algebras are isomorphic;
\item if the eigenvalues not coincide then such the algebras are not isomorphic.
\end{enumerate}
\end{algorithm}

This is really so, because any real symmetric spin-tensor can be transformed to the diagonal form by orthogonal transformations. In the other hand, the orthogonal transformation of the spinor space generates the orthogonal transformation of the base space, and hence this leads to the isomorphic algebra. Thus, really in order to compare the two algebras for n=8, it is sufficient to compare the eigenvalues of the controlling spin-tensors $\theta^{CD}$.

\begin{theorem}
The classification of metric hypercomplex group alternative-elastic algebras for n=8 can be reduced to the classification based on the eigenvalues of the symmetric controlling spin-tensor.
\end{theorem}

Thus, all is made necessary for a technical realization.

\begin{example} Technical realization is given in Appendix.
\begin{enumerate}
\item This article contains the file ''octonion.pas'' (by the operator $\backslash\ input\{octonion.pas\}$) being a programming unit adapted to the LaTex (LaTex version of this article is on arXiv) for the Delphi. At the same time, this file is Appendix to this article. You must create a project with this ''unit octonion'' and put on the form: Button1: TButton, ComboBox1: TComboBox, StringGrid1: TStringGrid (the lines 22-24).
\item At the lines 100-119,  the structural constants $\eta_{ij}{}^k$ for $\mathbb R^8$ ($i,j,k,...=\overline{1,8}$) are initialized for Example \ref{ex3}.
\item At the lines 122-141,  the structural constants $\eta_{ij}{}^k$ for $\mathbb R^8$ are initialized for Example \ref{ex4}.
\item At the lines 144-163,  the structural constants $\eta_{ij}{}^k$ for $\mathbb R^8$ are initialized for Example \ref{ex5}.
\item At the lines 166-185,  the structural constants $\eta_{ij}{}^k$ for $\mathbb R^8$ are initialized for Example \ref{ex6}.
\item At the lines 188-207,  the structural constants $\eta_{ij}{}^k$ for $\mathbb R^8$ are initialized for Example \ref{ex8}.
\item At the lines 210-227,  the connecting operators $\sqrt{2}\eta_i{}^{AB}$ for $\mathbb R^8$ ($i,A,B=\overline{1,8}$) are constructed.
\item At the lines 230-251,  the transition to the new basis is constructed.
\item At the lines 252-281,  the symmetric spin-tensor $\theta^{CD}$ is constructed.
\item At the lines 284-294,  the symmetric spin-tensor is outputted.
\item At the lines 295-324,  the structural constants are constructed on the symmetric spin-tensor with the help of the reverse motion.
\item At the lines 325-338,  the symmetric spin-tensor is outputted into the file.
\item System characteristics of the computer on which the program is tested: \\ HP Pavilion dv7-6b50er i3-2330M/4096/500/Radeon HD6770 2Gb/Win7 HP64
\item Run time <1 sec.
\end{enumerate}
\end{example}

\newpage
\hspace{12cm}\textbf{Appendix.}\\ \tiny
\lstset{basicstyle=\tiny}
\input{octonion.pas}
\newpage
\References{
\bibitem{Albert1}
A.A. Albert. Quadratic Forms Permitting Composition. Ann. of Math. 1942, 43, 161-177.
\bibitem{Andreev0}
К.В. Андреев  [K.V. Andreev]. О спинорном формализме при четной размерности базового пространства [O spinornom formalizme pri chetno\u\i razmernosti basovogo prostranstva]. ВИНИТИ - 298-B-11 [VINITI-298-V-11], июнь 2011 [Jun 2011], 138 с [138pp].  [in Russian: On the spinor formalism for the base space of even dimension. Paper deponed on Jun 16, 2011 at VINITI (Moscow), ref. \No 298-V 11]
\bibitem{Andreev1}
K.V. Andreev. On the spinor formalism for even n. \href{http://arxiv.org/abs/1110.4737}{arXiv:1110.4737v3} [math-ph].
[with the Russian edition: К.В. Андреев. О спинорном формализме при четной размерности базового пространства.].
\bibitem{Andreev2}
К. В. Андреев [K.V. Andreev]: Спинорный формализм и геометрия шестимерных римановых пространств [Spinorny\u\i\ formalizm i geometriya shestimernykh rimanovykh prostranstv]. Кандидатская диссертация [Kandidatskaya dissertatsiya], Уфа [Ufa], 1997, [in Russian: Spinor formalism and the geometry of six-dimensional Riemannian spaces. Ph. D. Thesis], \href{http://arxiv.org/abs/1204.0194}{arXiv:1204.0194v1} (with the Russian edition).
\bibitem{Andreev3}
K.V. Andreev. On the metric hypercomplex group alternative-elastic algebras for n mod 8 = 0. \href{http://arxiv.org/abs/1202.0941}{arXiv:1202.0941v1} [math-ph]
\bibitem{Baez1}
John C. Baez The Octonions. Bull. Amer. Math. Soc. 39 (2002), 145-205, \href{http://arxiv.org/pdf/math/0105155v4.pdf}{arXiv:math.RA/0105155v4}.
[Баэз Джон С. Октонионы.// Гиперкомплексные числа в геометрии и физике. \No 1(5), Vol. 3 (2006), c.120-177].
\bibitem{Moreno1}
R. Guillermo Moreno. The zero divisors of the Cayley-Dickson algebras over the real numbers. \href{http://arxiv.org/pdf/q-alg/9710013}{arXiv:q-alg/9710013v1} [math.QA]
\bibitem{Postnicov1}
М.М. Постников. [M.M. Postnikov] Группы и алгебры Ли [Gruppy i algebry Li]. Наука [Nauka], Москва [Moskva], 1986. English translation: M. Postnikov: Lie Groups and Lie Algebras. Lectures in Geometry, Semester 5. Mir, Moscow, 1986; URSS Publishing, Moscow, 1994. The main ideas of the hypercomplex number constraction on the base of the Bott periodicity are given in the lectures 13-16.
}
\end{document}